\documentclass{moriond}
\usepackage{hyperref}

\def\be{\begin{equation}}
\def\ee{\end{equation}}
\def\bea{\begin{eqnarray}}
\def\eea{\end{eqnarray}}

\newcommand{\Photo}{}

\begin{document}
\vspace*{4cm}
\title{Overview of Applications of Quantum Computing in QCD}

\author{Germ\'an Rodrigo}

\address{Instituto de F\'{\i}sica Corpuscular, Universitat de Val\`{e}ncia -- Consejo Superior de Investigaciones Cient\'{\i}ficas, Parc Cient\'{\i}fic, E-46980 Paterna, Valencia, Spain}

\maketitle

\abstract{
Quantum computing has emerged as a promising framework for addressing computationally demanding problems in collider physics. In recent years, a growing number of quantum algorithms have been proposed for applications ranging from event generation and parton shower simulation to the evaluation of scattering amplitudes, loop and phase-space integration, and optimization problems relevant to experimental analysis. We provide a concise overview of the main ideas behind these developments, with emphasis on the potential advantages of quantum approaches in comparison with classical methods, as well as on the current limitations imposed by noisy intermediate-scale quantum hardware.}

\section{The Precision Frontier}

As the era of CERN’s High-Luminosity Large Hadron Collider (HL-LHC) approaches, the particle physics community faces a crucial computational challenge. The HL-LHC is expected to collect approximately 10 times more data than the LHC has collected to date, including the Run 3, which concluded on 29 June 2026, which will enable measurements with unprecedented statistical precision and, at the same time, will reveal an emerging {\it precision gap} between experimental sensitivity and theoretical predictions. For example, projected experimental uncertainties for Higgs boson couplings~\cite{Cepeda:2019klc}, such as those to photons, $W$, and $Z$ bosons, are anticipated to reach 1.5\%–1.8\%, where, assuming a factor 2 improvement, the theoretical uncertainty for the $Z$ coupling is approximately 1.2\%. The precision gap becomes even more pronounced for other key couplings: the extrapolated theoretical uncertainties associated with the gluon and top quark couplings are estimated at 2.1\% and 3.1\%, respectively, thereby constituting the dominant contributions to the overall uncertainty.

The implications of this theoretical bottleneck are considerable. Without significant improvements to reduce theoretical uncertainties below the current 2\%–4\% range, it will not be possible to fully exploit the HL-LHC’s potential for discovery. In this context, the limiting factor is no longer the collider’s luminosity, but rather the precision of  theoretical calculations, particularly in Quantum Chromodynamics (QCD). Addressing this challenge requires a fundamental shift in computational strategy, aligning theoretical methods with the intrinsically quantum nature of the underlying physical system.

\section{High-Energy Colliders are Quantum Machines}

A fundamental principle of High-Energy Physics (HEP) is that particle interactions probed at collider experiments are described by Quantum Field Theory~(QFT). Given that Quantum Mechanics~(QM) constitutes a central component of QFT, with QM $\subset$ QFT, it is natural to expect that the simulation of collider phenomena in classical computers may entail intrinsic inefficiencies, whilst the transition to quantum-based simulations 
will more faithfully reflect the quantum nature of the physical processes under study. In this regard, it echoes Feynman’s well-known motivation for Quantum Computing (QC)~\cite{Feynman:1981tf}: “Nature isn’t classical, dammit, and if you want to make a simulation of nature, you better make it quantum.” 
From this perspective, the LHC may be regarded as the {\it largest quantum machine} ever constructed, and the Future Circular Collider (FCC)~\cite{FCC:2025lpp} represents the next generation of such quantum systems. This interpretation is further reinforced by the historical connection between CERN and quantum information science, with CERN the place {\it where the Bell inequalities were born}~\cite{Bell:1964kc}~\footnote{Published during a period of leave of absence from CERN.}. 

The three fundamental principles of QM are {\it superposition}, {\it entanglement}, and {\it interference}. Superposition is the ability of a system to be simultaneously in multiple states until measured. Entanglement is the ability of systems to correlate their states across distances. Einstein called this "spooky action at a distance". Interference is the ability to combine probability amplitudes to enhance or suppress an outcome. It must be emphasized that QC differs substantially from a fast parallel computing. Although a quantum system simultaneously explores the entire probability space of potential solutions through superposition, the final measurement collapses the wave function to yield just one specific outcome. It is also important to stress that for QC to offer a computational advantage over classical approaches, quantum algorithms must leverage all three QM fundamental  principles in an integrated manner. 

In addition, QC should not merely be regarded as a means of accelerating computations; it also introduces novel conceptual frameworks that can inspire improvements in classical algorithms by recasting the same problem from a different perspective. In this broader sense, {\it quantum-inspired} methods may enhance classical approaches. Another important concept is {\it quantum utility}. It denotes the regime in which QC provides value by enabling the inclusion of physical effects that are neglected in classical approximations, thereby leading to more accurate theoretical predictions. This is particularly relevant in collider physics, where quantum algorithms may provide more faithful theoretical predictions by treating interference, entanglement, and other genuinely quantum features that are difficult to encode in standard classical approximations.

\section{Noisy Intermediate-Scale Quantum versus Fault-Tolerant Eras}

The transition from theoretical QC to practical applications demands  significant advances in quantum hardware. We are currently operating in the Noisy Intermediate-Scale Quantum~(NISQ)~\cite{Preskill:2018jim} era, with a roadmap toward Fault-Tolerant Quantum Computing (FTQC). NISQ devices typically comprise thousands of physical qubits; however, these qubits remain noisy and exhibit short coherence times. For instance, superconducting qubits have coherence times of ${\cal O}(100~\mu s)$, whereas trapped-ion systems can reach ${\cal O}(10~s)$ but at the cost of slower gate operations. FTQC is expected to rely on ${\cal O}(100)$ error-corrected logical qubits, and some vendors claim that such capability may be achievable within approximately five years. It should also be noted that noiseless classical quantum simulators can currently handle only about ${\cal O}(30)$ qubits.

There are two main paradigms in QC: gate-based and quantum annealing. Gate-based QC employs universal logic gates to execute general-purpose algorithms, whereas quantum annealing is designed for energy minimization, that is, for finding the global minimum of a Hamiltonian or cost function. Gate-based QC is more flexible, while quantum annealers currently offer greater qubit scalability, making them particularly well suited to optimization tasks.

\section{Quantum Algorithms in High-Energy Physics}

As in many other fields, QC has attracted an increasing attention in particle physics in recent years. In collider physics, the most studied applications include parton densities~\cite{Perez-Salinas:2020nem,Kang:2025xpz}, parton showers~\cite{Bauer:2019qxa,Bepari:2020xqi,Bepari:2021kwv,Rouxinol:2026suk}, fragmentation functions~\cite{deLejarza:2025upd}, tree-level helicity amplitudes~\cite{Bepari:2020xqi,Bashore:2025uwb,Haddad:2026ncx}, colour algebra~\cite{Chawdhry:2023jks,Chawdhry:2025iuz}, causal configurations of multiloop Feynman diagrams~\cite{Ramirez-Uribe:2021ubp,Clemente:2022nll,Ramirez-Uribe:2024wua,Ochoa-Oregon:2025opz}, 
integration of one-dimensional and multidimensional functions~\cite{Herbert:2021xgs,Agliardi:2022ghn,Cruz-Martinez:2023vgs,deLejarza:2023qxk,Pyretzidis:2025stx,Williams:2025hza}, loops~\cite{deLejarza:2024pgk,Pyretzidis:2025stx} and decay rates from multiloop vacuum amplitudes~\cite{Ramirez-Uribe:2024rjg,LTD:2024yrb,deLejarza:2024scm}, jet physics in vacuum~\cite{Wei:2019rqy,Pires:2020urc,Felser:2020mka,deLejarza:2022bwc,Delgado:2022snu,Okawa:2024goh} or in a medium~\cite{Barata:2021yri,Barata:2023clv,Barata:2026icn}, track reconstruction~\cite{Magano:2021jzd,Duckett:2022ccc,Nicotra:2023rmn,Okawa:2024eof,Chiotopoulos:2026jas}, anomaly detection~\cite{Ngairangbam:2021yma,Belis:2023atb,Schuhmacher:2023pro,Laino:2025ksm}, and other data-intensive tasks relevant to the HL-LHC and future colliders. Event generation, viewed as the sampling of quantum circuits, is perhaps the most natural application of QC in collider physics, offering a compelling vision for the future of precision event simulation. 

A standout development is the Variational Quantum Machine Learning (QML) workflow. These models typically require far fewer parameters~\cite{Pyretzidis:2025stx} than classical neural networks to achieve high expressivity, and provide a better mapping of correlations. In the NISQ era, this efficiency represents a viable path to achieving meaningful results on current hardware, allowing us to identify subtle patterns in LHC data that classical architectures might miss.

For perturbative calculations, quantum algorithms have been proposed to accelerate resource-intensive tasks in high-order expansions, such as Monte Carlo integration and sampling that arise in amplitude and cross-section computations. The motivation is not that a quantum computer directly replaces a perturbative calculation, but that it may reduce the cost of bottlenecks inside the workflow, especially when many integrals, samples, or configurations must be evaluated.

A practical theme is that near-term quantum devices are still noisy and limited, so the field currently focuses on benchmark problems and hybrid quantum-classical methods rather than full-scale production use. The most credible short-term opportunities are therefore in small but representative collider applications and in testing whether quantum methods improve specific computational methods used high-energy physics.

\section{Conclusions}

High-energy colliders are inherently quantum systems, with the LHC representing the largest quantum machine built to date and the FCC to become the next. This makes collider physics a particularly well-matched domain for QC approaches. QC may offer advantages for problems, or subproblems, whose computational complexity grows rapidly, such that they become untractable with classical computers, or where correlations play a central role, since such cases can potentially benefit from the QM principles of superposition, entanglement, and interference. At present, however, classical methods remain more effective in practice, although leading vendors anticipate the availability of 50–100 logical qubits within five years.

Several promising applications have already been developed in the field of collider physics, which could become scalable once quantum hardware with sufficient capacity becomes available. In the long term, an appealing objective is the development of a fully-fledged quantum event generator that reaches theoretical predictions at high perturbative orders, overcoming the bottlenecks that limit the current state of the art.

\section*{Acknowledgments}

This work is supported by the Spanish Government and ERDF/EU - Agencia Estatal de Investigaci\'on MCIN/AEI/10.13039/501100011033,  Grants No. PID2023-146220NB-I00, No. EUR2025-164820, and No. CEX2023-001292-S; and Quantum Spain (Ministry of Economic Affairs and Digital Transformation,  NextGenerationEU).

\section*{References}


\end{document}